\documentclass{ws-mpla}

\usepackage[mathscr]{eucal}
\usepackage{overcite}

\providecommand{\dalembert}{\square}%
\providecommand{\lagrange}{\mathscr{L}}%
\providecommand{\del}{\partial}%
\providecommand{\I}{i}

\begin{document}

\markboth{Roberto Baginski B. Santos}
{Plasma-like Vacuum in Podolsky Regularized
Classical Electrodynamics}

\catchline{}{}{}{}{}

\title{PLASMA-LIKE VACUUM IN PODOLSKY REGULARIZED CLASSICAL ELECTRODYNAMICS}

\author{\footnotesize ROBERTO BAGINSKI B. SANTOS}

\address{Departamento de F\'{\i}sica, Centro Universit\'{a}rio da FEI\\ Avenida Humberto de A. C. Branco 3972\\ S\~{a}o Bernardo do Campo, SP 09850-901, Brazil\\
rsantos@fei.edu.br}

\maketitle

\pub{Received (Day Month Year)}{Revised (Day Month Year)}

\begin{abstract}
We analyze wave propagation in the vacuum of Podolsky regularized electrodynamics. Two kinds of waves were found in the theory: the traditional non-dispersive waves of Maxwell electrodynamics, and a dispersive wave reminiscent of wave propagation in a collisionless plasma. Charged particle concentration was determined, and found to be huge in this vacuum. We interpret the results in terms of vacuum polarization effects induced in an otherwise classical theory.

\end{abstract}

\ccode{PACS Nos.: 03.50.De, 03.65.Sq, 12.90.+b}

\section{Introduction}
\label{sec:intro}

Classical electrodynamics has been extremely successful for the past 150 years or more. However, whenever it dealt with point charged particles, the results were disappointing: a divergent electromagnetic energy, the infamous $4/3$ problem of the electromagnetic mass in the Abraham--Lorentz theory, and the runaway solutions of the classical Lorentz--Dirac equation of motion are all symptoms of a deeper maladie. In our view, there is no really satisfactory way to solve these issues entirely within the classical context. All of these problems occur in a very small length scale, in which classical electrodynamics is not supposed to work properly. Therefore, any solution to these problems will have to involve contributions from processes that take place in the quantum realm.

Despite its impressive record, quantum electrodynamics is also plagued by the same type of problems encountered in many linear quantum theories~\cite{Norton+Watson:1959}. In fact, any quantum theory will feature runaway solutions if its classical counterpart also have it~\cite{Coleman:1962}. Again, the problem is that the introduction of point particles in a theory leads us directly into length, time, and energy scales in which strong, weak, and even gravitational phenomena are expected to play a significant role. Owing to regularization and renormalization techniques, which summed up all these high energy contributions in an effective way, accurate results were achieved in quantum electrodynamics.

In the context of the quantum theory of non-relativistic electrons, it was found that the interaction of a point electron with its own electromagnetic field induces an effective cutoff of the order of the electron reduced Compton wavelength $\lambda_\mathrm{C}=\hbar/mc\approx 386\;\mathrm{fm}$~\cite{Moniz+Sharp:1974, Moniz+Sharp:1977,
Levine+Moniz+Sharp:1977}. This cutoff owes its existence to \emph{Zitterbewegung}, the jittery motion caused by the never ending creation and annihilation of virtual electron--positron pairs around the point particle, effectively spreading its charge over a region of length comparable to $\lambda_\mathrm{C}$.

Therefore, classical electrodynamics is a theory valid at a length scale in which quantum phenomena are not very important, a few dozen Bohr radius, for instance. However, in order to describe some of the physical phenomena that take place in a length scale comparable to the electron Compton wavelength, we must extend classical electrodynamics, treating it as an effective theory in which a cutoff owes its existence to quantum phenomena at small distance. In sections~\ref{sec:Podolsky_electrodynamics} and~\ref{sec:field_and_potential}, we present a brief review of some aspects of the Podolsky regularized electrodynamics in a classical context, in which a second-order derivative term that introduces a cutoff $\ell$ to the electromagnetic interaction is added to the Maxwell lagrangian density in order to allow us to describe a range of phenomena in which vacuum polarization is important.

\section{Podolsky regularized electrodynamics}
\label{sec:Podolsky_electrodynamics}

Classical electrodynamics is a linear theory. Although interesting, attempts to formulate a nonlinear electrodynamics have not gained enough traction~\cite{Proca:1930,Born+Infeld:1934a}. In order to preserve the linear structure of classical electrodynamics, and still allow for a cutoff $\ell$ into the theory in a Lorentz and gauge invariant way, a term involving second order derivatives of the electromagnetic potential $A$ may be introduced in the lagrangian density for the electromagnetic field. In this case, the lagrangian density reads
\begin{equation}\label{eq:lagrangian}
 \lagrange=-\frac{1}{16\pi}F^{\mu\nu}F_{\mu\nu}
 -\frac{\ell^2}{8\pi}\,\del^\alpha F_{\mu\alpha}\del_\beta
 F^{\mu\beta} + \frac{1}{c}\,j^\mu A_\mu,
\end{equation}
where, as usual, $F^{\mu\nu}=\del^\mu A^\nu - \del^\nu A^\mu$ are the components of the electromagnetic field tensor $F$, and $j$ is the current.

The middle extra term was proposed long ago in an effort to regularize quantum electrodynamics~\cite{Podolsky:1942, Podolsky+Kikuchi:1944, Podolsky+Kikuchi:1945, Podolsky+Schwed:1948}. At about the same time, a number of equivalent proposals were made~\cite{Bopp:1940, Lande:1941, Lande+Thomas:1941, Bopp:1943, Lande+Thomas:1944, Feynman:1948}. Recently, it was shown that Podolsky lagrangian is the only linear second-order gauge-invariant generalization of Maxwell electrodynamics~\cite{Cuzinatto+deMelo+Pompeia:2007}.

Regarding quantum electrodynamics, Podolsky proposal to generalize electrodynamics is akin to Pauli--Villars regularization procedure~\cite{Pauli+Villars:1949}. In the Pauli-Villars regularization of the electron self-energy, an extra term is introduced in the lagrangian density, corresponding to a heavy auxiliary particle. The mass of this particle is related to a cutoff $\ell$, which tames the infinities of the theory, by $M=\hbar/c\ell$. As the cutoff goes to zero, the mass of the auxiliary particle tends to infinity and disappears from the theory. Nowadays, higher order derivatives appears in attempts to regularize various gauge theories~\cite{Slavnov:1971,
Slavnov:1972, Slavnov:1978, Namazie:1980, West:1986, Polyakov:1986, Faddeev+Slavnov:1991, Barcelos-Neto+Galvao+Natividade:1991, Rubakov:2002, Grinstein+OConnell+Wise:2008, Cuzinatto+deMelo+Medeiros+Pompeia:2008}.

The good ultraviolet behavior of Podolsky quantum electrodynamics comes at the cost of introducing a non-tachyonic ghost in the theory~\cite{Accioly+Mukai:1997}. Therefore, Podolsky quantum electrodynamics may be viewed as an effective field theory as this kind of ghost may lead to non-unitary evolution in a quantum theory~\cite{Matthews:1949,Bloch:1952,Sakurai:1967}. Despite that, it was pointed out that magnetic monopoles and massive photons may coexist in Podolsky quantum electrodynamics~\cite{Fonseca+Paredes:2010}. In fact, this coexistence is not ruled out by the analysis performed in finite-range electrodynamics~\cite{Ignatiev+Joshi:1996} owing to the fact that Podolsky quantum electrodynamics is a truly long-range electrodynamics with a massless excitation accompanied by a massive one. However, it may be argued that the massive photon of Podolsky quantum electrodynamics is not physically sound~\cite{Kruglov:2010}.

However, when dealing with Podolsky regularized electrodynamics as an effective theory aiming at introducing some quantum effects in a otherwise classical realm, these troubles are avoided. At the same time, we may achieve a more vivid description of the physical phenomena. In Podolsky regularized classical electrodynamics, it was possible to solve the infamous $4/3$-problem~\cite{Frenkel:1996}, and to eliminate runaway solutions from the Lorentz--Dirac equation of motion~\cite{Frenkel+Santos:1999}. Requiring that the correction to the hydrogen ground state energy be smaller than the relative experimental uncertainty, it was shown that the cutoff $\ell\leq 5.56\;\mathrm{fm}$~\cite{Cuzinatto+deMelo+Medeiros+Pompeia:2009}. Hence, the cutoff length scale is well within the range of quantum phenomena such as pair creation and annihilation, and \emph{Zitterbewegung}.

\section{Field and potential equations in Podolsky regularized electrodynamics}
\label{sec:field_and_potential}

In Podolsky electrodynamics, the potential $A$ obeys the equations
\begin{equation}\label{eq:069}
 \bigl(1-\ell^2\dalembert\bigr)\bigl(\dalembert A^\nu-\del^\nu\del_\mu
 A^\mu\bigr)=-\frac{4\pi}{c}j^\nu,
\end{equation}
where $\dalembert=\del^\mu\del_\mu$ is the usual D'Alembert differential operator. These equations for the potential lead to electromagnetic field equations
\begin{equation}\label{eq:070}
 \bigl(1-\ell^2\dalembert\bigr)\del_\mu F^{\mu\nu} = -\frac{4\pi}{c}j^\nu
\end{equation}
that are of fourth order in the field $F$ in contrast to the usual Maxwell equations. As $F^{\mu\nu} = \del^\mu A^\nu
- \del^\nu A^\mu$ are still the components of an antisymmetric tensor, Bianchi identities
\begin{equation}\label{eq:072}
 \del_\lambda F_{\mu\nu}+\del_\mu F_{\nu\lambda}+\del_\nu
 F_{\lambda\mu} = 0
\end{equation}
still hold. We see that the regulator modifies only Coulomb--Gauss and Amp\`{e}re--Maxwell laws, altering the relationship between the electromagnetic field $F$ and its sources $j$ only at small distances. In order to see that,  consider a point particle at rest. It is easy to show that the electric potential $\Phi(\vec{r}) = \bigl(1-\exp(-r/\ell)\bigr)e/r$ in regularized electrodynamics
tends to the usual $\Phi(\vec{r})=e/r$ at large distances $r$ from the particle, and to the finite $\Phi(\vec{r})=e/\ell$ at small distances. Therefore, Podolsky regularized electrodynamics alters the behavior of the field only at small distances from its source. At large distances, we recover Maxwell electrodynamics.

To see that this modification is generally effective only at small distances, we write Eq.~\ref{eq:069} in the Lorenz gauge,
\begin{equation}\label{eq:073}
 \bigl(1-\ell^2\dalembert\bigr)\dalembert
 A^\mu=-\frac{4\pi}{c}j^\mu,
\end{equation}
and represent the potential $A(x)$ as as a linear combination of propagating plane waves
\begin{equation}\label{eq:074}
 A^\mu(x)=\int\widetilde{A}^\mu(k)\exp(\I k^\nu x_\nu)\,d^4k
\end{equation}
while the current density $j(x)$ is expanded accordingly as
\begin{equation}\label{eq:075}
 j^\mu(x)=\int\widetilde{j}^\mu(k)\exp(\I k^\nu x_\nu)\,d^4k.
\end{equation}
It is, then, straightforward to show that
\begin{equation}\label{eq:076}
 -\int k^2\bigl(1+\ell^2k^2\bigr)\widetilde{A}^\mu(k) \exp(\I k^\nu x_\nu)\,d^4k
 =-\frac{4\pi}{c}\int\widetilde{j}^\mu(k)\exp(\I k^\nu x_\nu)\,d^4k,
\end{equation}
where $k^2=k^\mu k_\mu$. From Eq.~\ref{eq:076}, we easily arrive at
\begin{equation}\label{eq:077}
 \widetilde{A}^\mu(k)=\frac{4\pi}{c}\,\frac{1}{k^2\bigl(1+\ell^2k^2\bigr)}\,\widetilde{j}^\mu(k),
\end{equation}
 and discover that $\widetilde{A}(k)$ is reduced to the usual Maxwell expression
\begin{equation}\label{eq:077Maxwell}
 \widetilde{A}_\mathrm{M}^\mu(k)=\frac{4\pi}{c}\,\frac{1}{k^2}\,\widetilde{j}^\mu(k),
\end{equation}
at large distances (small $k$) while $\widetilde{A}(k)$ tends to a distinctive $1/k^4$ form,
\begin{equation}\label{eq:077-1}
 \widetilde{A}^\mu(k)=\frac{4\pi}{c}\,\frac{1}{\ell^2 k^4}\,\widetilde{j}^\mu(k),
\end{equation}
at small distances (large $k$).

\section{Wave propagation in Podolsky regularized electrodynamics}
\label{sec:wave_propagation}

These results hint that the regulator also affects the free space wave propagation, allowing for the coexistence of propagating and evanescent modes even in vacuum. Setting $j=0$ in Eq.~\ref{eq:076}, we find
\begin{equation}\label{eq:087}
 k^2\bigl(1+\ell^2k^2\bigr)=0,
\end{equation}
As $k=(\omega/c,\vec{\kappa})$, we may rewrite Eq.~\ref{eq:087} as
\begin{equation}\label{eq:088}
 \biggl(-\frac{\omega^2}{c^2}+\kappa^2\biggr)
 \Biggl(1+\ell^2\biggl(-\frac{\omega^2}{c^2}+\kappa^2\biggr)\Biggr)=0,
\end{equation}
from which we can derive two different dispersion relations for the propagation of an electromagnetic wave in free space. The first dispersion relation is the familiar $\kappa=\omega/c$ of Maxwell electrodynamics. This linear dispersion relation corresponds to non-dispersive wave propagation with phase velocity $c$.

The second dispersion relation,
\begin{equation}\label{eq:089}
 \kappa^2=\frac{\omega^2}{c^2}-\frac{1}{\ell^2}.
\end{equation}
may be more familiar in the form
\begin{equation}\label{eq:090}
 c\kappa=\sqrt{\omega^2-\omega_\mathrm{p}^2},
\end{equation}
where $\omega_\mathrm{p}=c/\ell\approx 1.02\;\mathrm{MeV}/\hbar$ if $\ell\approx\lambda_\mathrm{C}/2$. Eq.~\ref{eq:090} describes electromagnetic wave propagation through a colisionless plasma~\cite{Lieberman+Lichtenberg:1994,Jackson:1998} with a very small attenuation length $\delta= c/\omega_\mathrm{p}=\ell$.

\begin{figure}[htb]
 \centering
 \includegraphics{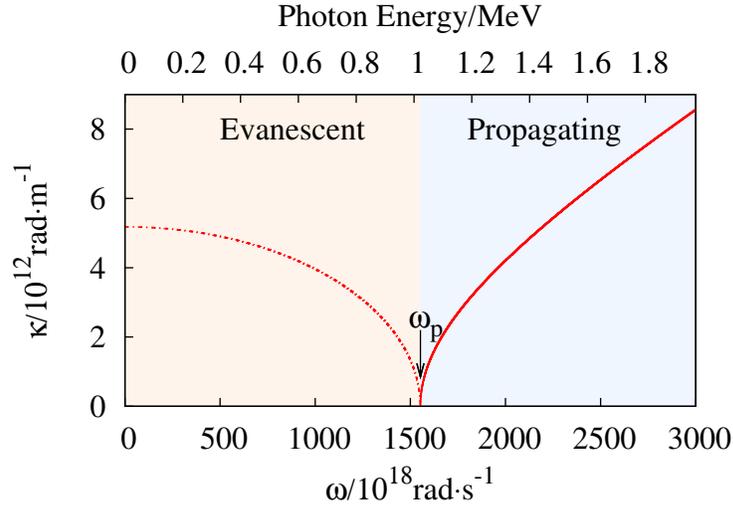}
 \caption{Dispersion relation for free-space dispersive wave propagation in Podolsky regularized electrodynamics. Continuous line corresponds to propagating modes, while the dashed line is the imaginary part of $\kappa$ associated with evanescent modes.}
 \label{fig:1}
\end{figure}

In fact, from Eq.~\ref{eq:089}, we see that the wavenumber $\kappa$ is real only if $|\omega|\geqslant c/\ell$. Otherwise, $\kappa$ is a pure imaginary number, leading to evanescent wave modes as shown in figure~\ref{fig:1}.

We can determine phase and group velocity for the dispersive modes. While phase velocity is determined by the relation
\begin{equation}\label{eq:092}
 v_\mathrm{p}=\frac{\omega}{\kappa},
\end{equation}
group velocity is determined by
\begin{equation}\label{eq:093}
 v_\mathrm{g}=\frac{\del\omega}{\del
 \kappa}=c^2\frac{\kappa}{\omega}.
\end{equation}
As suggested by the plasma-like vacuum analogy we are pursuing, $v_\mathrm{p}>c$ and $v_\mathrm{g}<c$ for the propagating modes, for which $\kappa$ is real. On the other hand, $v_\mathrm{p}<c$ and $v_\mathrm{g}>c$ for the evanescent modes, for which $\kappa$ is imaginary.

Charged particle concentration in this plasma-like vacuum is given by
\begin{equation}\label{eq:concentration}
 n=\frac{m\omega_\mathrm{p}^2}{4\pi e^2}=\frac{mc^2}{4\pi e^2\ell^2}\approx 760\times 10^{30}\,\mathrm{cm}^{-3},
\end{equation}
if $\ell\approx\lambda_\mathrm{C}/2$. This huge particle density, corresponding to approximately $185$ charged particles popping in and out of existence in a sphere of radius $\lambda_\mathrm{C}$, may be interpreted as the cause of the electron jittery motion induced by pair creation and annihilation.

In a quantum theory, these results were interpreted as a sign of the existence of two excitations in Podolsky electrodynamics~\cite{Podolsky+Kikuchi:1944}, corresponding to the two kinds of wave we have found: a massless photon, and a massive neutral boson with mass in the range of the $Z$ boson~\cite{Frenkel+Santos:1999,Accioly+Scatena:2010}.

\section{Conclusions}
\label{sec:conclusions}

We analyzed wave propagation in the vacuum of Podolsky regularized electrodynamics, discovering two different waves: the usual non-dispersive wave with $\kappa=\omega/c$, and a dispersive wave with a propagating mode for high frequency ($\omega>\omega_\mathrm{p}$), and an evanescent mode for low frequency ($\omega<\omega_\mathrm{p}$). In a classical effective theory framework, we interpret this result as arising from a plasma-like behavior of the vacuum induced by quantum vacuum polarization. While high-energy photons ($\hbar\omega>\hbar\omega_\mathrm{p}$) may produce real electron-positron pairs, low-energy photons ($\hbar\omega<\hbar\omega_\mathrm{p}$) can only yield virtual pairs. These virtual particles, living in an otherwise classical vacuum, act like a plasma, disturbing wave propagation. Therefore, Podolsky regularized electrodynamics inserts some features of pair creation and annihilation into the classical domain. Interaction of free-space electromagnetic waves with these virtual particles is equivalent to wave propagation in an effective medium, the classical vacuum plus quantum vacuum polarization, that behaves like a plasma.


\begin{thebibliography}{42}
\bibitem{Norton+Watson:1959}
R.~E. Norton and W.~K.~R. Watson, \emph{Phys. Rev.} \textbf{116}, 1597 (1959).
\bibitem{Coleman:1962}
S.~Coleman, \emph{Phys. Rev.} \textbf{125}, 1422 (1962).
\bibitem{Moniz+Sharp:1974}
E.~J. Moniz, D.~H. Sharp, \emph{Phys. Rev. D} \textbf{10}, 1133 (1974).
\bibitem{Moniz+Sharp:1977}
E.~J. Moniz, D.~H. Sharp, \emph{Phys. Rev. D} \textbf{15}, 2850 (1977).
\bibitem{Levine+Moniz+Sharp:1977}
H.~Levine, E.~J. Moniz, D.~H. Sharp, \emph{Am. J. Phys.} \textbf{45}, 75 (1977).
\bibitem{Proca:1930}
A.~Proca, \emph{Compt. Rend.} \textbf{190}, 1377 (1930).
\bibitem{Born+Infeld:1934a}
M.~Born, L.~Infeld, \emph{Proc. Roy. Soc. A} \textbf{144}, 425 (1934).
\bibitem{Podolsky:1942}
B.~Podolsky, \emph{Phys. Rev.} \textbf{62}, 68 (1942).
\bibitem{Podolsky+Kikuchi:1944}
B.~Podolsky, C.~Kikuchi, \emph{Phys. Rev.} \textbf{65}, 228 (1944).
\bibitem{Podolsky+Kikuchi:1945}
B.~Podolsky, C.~Kikuchi, \emph{Phys. Rev.} \textbf{67}, 184 (1945).
\bibitem{Podolsky+Schwed:1948}
B.~Podolsky, P.~Schwed, \emph{Rev. Mod. Phys.} \textbf{20}, 40 (1948).
\bibitem{Bopp:1940}
F.~Bopp, \emph{Ann. Physik} \textbf{38}, 345 (1940).
\bibitem{Lande:1941}
A.~Land\'{e}, \emph{Phys. Rev.} \textbf{60}, 121 (1941).
\bibitem{Lande+Thomas:1941}
A.~Land\'{e}, L.~H. Thomas, \emph{Phys. Rev.} \textbf{60}, 514 (1941).
\bibitem{Bopp:1943}
F.~Bopp, \emph{Ann. Phys.} \textbf{42}, 573 (1943).
\bibitem{Lande+Thomas:1944}
A.~Land\'{e}, L.~H. Thomas, \emph{Phys. Rev.} \textbf{65}, 175 (1944).
\bibitem{Feynman:1948}
R.~P. Feynman, \emph{Phys. Rev.} \textbf{74}, 939 (1948).
\bibitem{Cuzinatto+deMelo+Pompeia:2007}
R.~R. Cuzinatto, C.~A.~M. de~Melo, P.~J. Pompeia, \emph{Ann. Phys.} \textbf{322}, 1211 (2007).
\bibitem{Pauli+Villars:1949}
W.~Pauli, F.~Villars, \emph{Rev. Mod. Phys.} \textbf{21}, 434 (1949).
\bibitem{Slavnov:1971}
A.~A. Slavnov, \emph{Nucl. Phys. B} \textbf{31}, 301 (1971).
\bibitem{Slavnov:1972}
A.~A. Slavnov, \emph{Theor. Math. Phys.} \textbf{13}, 1064 (1972).
\bibitem{Slavnov:1978}
A.~A. Slavnov, \emph{Theor. Math. Phys.} \textbf{33}, 997 (1978).
\bibitem{Faddeev+Slavnov:1991}
L.~D. Faddeev, A.~A. Slavnov, \emph{Gauge Fields: An Introduction to Quantum Theory} 2nd. edition (Addison-Wesley, 1991).
\bibitem{Rubakov:2002}
V.~A. Rubakov, \emph{Classical Theory of Gauge Fields} (Princeton University Press, 2002).
\bibitem{Grinstein+OConnell+Wise:2008}
B.~Grinstein, D.~O'Connell, M.~B. Wise, \emph{Phys. Rev. D} \textbf{77}, 025012 (2008).
\bibitem{Namazie:1980}
M.~A. Namazie, \emph{J. Phys. A} \textbf{11}, 713 (1980).
\bibitem{West:1986}
P.~West, \emph{Nucl. Phys. B} \textbf{268}, 113 (1986).
\bibitem{Polyakov:1986}
A.~M. Polyakov, \emph{Nucl. Phys. B} \textbf{268}, 406 (1986).
\bibitem{Barcelos-Neto+Galvao+Natividade:1991}
J.~Barcelos-Neto, C.~A.~P. Galv\~{a}o, C.~P. Natividade, \emph{Z. Phys. C} \textbf{52}, 559 (1991).
\bibitem{Cuzinatto+deMelo+Medeiros+Pompeia:2008}
R.~R. Cuzinatto, C.~A.~M. de~Melo, L.~G. Medeiros, P.~J. Pompeia, \emph{Eur. Phys. J. C} \textbf{53}, 99 (2008).
\bibitem{Accioly+Mukai:1997}
A.~Accioly, H.~Mukai, \emph{N. Cim. B} \textbf{112}, 1061 (1997).
\bibitem{Bloch:1952}
C.~Bloch, \emph{Kgl. Danske Videnskab Selskab Mat.-Fys. Medd.} \textbf{27}, No. 8, 2 (1952).
\bibitem{Sakurai:1967}
J.~J. Sakurai, \emph{Advanced Quantum Mechanics} (Addison-Wesley, 1967).
\bibitem{Matthews:1949}
P.~T. Matthews, \emph{Math. Proc. Cambridge Phil. Soc.} \textbf{45}, 441 (1949).
\bibitem{Fonseca+Paredes:2010}
M.~V.~S. Fonseca, A.~V. Paredes, \emph{Braz. J. Phys.} \textbf{40}, 319 (2010).
\bibitem{Ignatiev+Joshi:1996}
A.~Y. Ignatiev, G.~C. Joshi, \emph{Phys. Rev. D} \textbf{53}, 984 (1996).
\bibitem{Kruglov:2010}
S.~I. Kruglov, \emph{J. Phys. A: Math. Theor} \textbf{43}, 245403 (2010).
\bibitem{Frenkel:1996}
J.~Frenkel, \emph{Phys. Rev. E} \textbf{54}, 5859 (1996).
\bibitem{Frenkel+Santos:1999}
J.~Frenkel, R.~B. Santos, \emph{Int. J. Mod. Phys. B} \textbf{13}, 315 (1999).
\bibitem{Cuzinatto+deMelo+Medeiros+Pompeia:2009}
R.~R. Cuzinatto, C.~A.~M. de~Melo, L.~G. Medeiros, P.~J. Pompeia, arXiv:0810.4106v2 [quant-ph] (2009).
\bibitem{Lieberman+Lichtenberg:1994}
M.~A. Lieberman, A.~Lichtenberg, \emph{Principles of Plasma Discharges and Materials Processing} (Wiley, 1994).
\bibitem{Jackson:1998}
J.~D. Jackson, \emph{Classical Electrodynamics} 3rd. edition (Wiley, 1998).
\bibitem{Accioly+Scatena:2010}
A.~Accioly, E.~Scatena, \emph{Mod. Phys. Lett. A} \textbf{25}, 1115 (2010).

\end{thebibliography}
\end{document}